 \documentclass[authoryear,preprint,review,12pt]{elsarticle}

\usepackage{graphicx}

\usepackage{amssymb}
\usepackage{epsf}
\usepackage{psfrag}
\usepackage{amsmath}
\usepackage{lineno}
\usepackage{numcompress}

\journal{J. Atm. Sol-Ter. Phys.,}

\begin{document}

\begin{frontmatter}



\title{Analysis on diurnal global geomagnetic variability under quiet-time conditions}


\author{Virginia Klausner\corref{cor}\fnref{footnote2}}
\address{DGE/CEA/National Institute for Space Research - phone: +55 12 32086819/fax: +55 12 32086810 - INPE 12227-010 S\~ao Jos\'e dos Campos, SP, Brazil.}
\cortext[cor]{Corresponding author}
\fntext[footnote2]{National Observatory - ON 20921-400, RJ, Brazil.}
\ead{virginia@dge.inpe.br}

\author{Margarete Oliveira Domingues}
\address{LAC/CTE/National Institute for Space Research - phone: +55 12 32086542/fax: +55 12 32086375 - INPE 12227-010 S\~ao Jos\'e dos Campos, SP, Brazil}
\ead{mo.domingues@lac.inpe.br}

\author{Odim Mendes Jr}
\address{DGE/CEA/National Institute for Space Research - phone: +55 12 32087854/fax: +55 12 32086810 - INPE 12227-010 S\~ao Jos\'e dos Campos, SP, Brazil}
\ead{odim@dge.inpe.br}

\author{Andres Reinaldo Rodriguez Papa\fnref{footnote5}}
\address{National Observatory - phone: +55 21 35049142/fax: +55 21 25807081 - ON 20921-400, RJ, Brazil}
\fntext[footnote5]{State University of Rio de Janeiro - UERJ 20550-900, RJ, Brazil}
\ead{papa@on.br}

\author{Peter Frick}
\address{Laboratory of Physical Hydrodynamics - phone: +7 342 2378322/fax: +7 342 2378487 - Institute of Continuous Media Mechanics, Perm, Russia}
\ead{frick@icmm.ru}



\begin{abstract}
This paper describes a methodology (or treatment) to establish a representative signal of the global magnetic diurnal variation based on a spatial distribution in both longitude and latitude of a set of magnetic stations as well as their magnetic behavior on a time basis.
For that, we apply the Principal Component Analysis (PCA) technique implemented using gapped wavelet transform and wavelet correlation.
The continuous gapped wavelet and the wavelet correlation techniques were used to describe the features of the magnetic variations at Vassouras (Brazil) and other $12$ magnetic stations spread around the terrestrial globe.
The aim of this paper is to reconstruct the original geomagnetic data series of the H-component taking into account only the diurnal variations with periods of $24$ hours on geomagnetically quiet days.
With the developed work, we advance a proposal to reconstruct the baseline for the quiet day variations (Sq) from the  PCA using the correlation wavelet method to determine the global variation of PCA first mode.
The results showed that this goal was reached and encourage other uses of this approach to different kinds of analysis.
\end{abstract}


\end{frontmatter}
\linenumbers

\section{Introduction}
\label{Introduction}

A substantial part of the energy carried by the solar wind can be transfered into the terrestrial magnetosphere and it is associated with the passage of southward directed interplanetary magnetic fields, Bs, by the Earth for sufficiently long intervals of time.
\Citet{Gonzalezetal:1994} discussed the energy transfer process as a conversion of the directed mechanical energy from the solar wind into magnetic energy stored in the magnetotail of Earth's magnetosphere and its reconversion into thermal mechanical energy in the plasma sheet, auroral particles, ring current, and Joule heating of the ionosphere.

The increase on the solar wind pressure is responsible for the energy injections and induces global effects in the magnetosphere called geomagnetic storms. 
The characteristic signature of geomagnetic storms can be described as a depression on the horizontal component of the Earth's magnetic field measured at low and middle latitude ground stations.
The decrease in the magnetic horizontal field component is due to an enhancement of the trapped magnetospheric particle population, consequently an enhanced ring of current. 
This perturbation of the H-component could last from several hours to several days \cite[as described by][]{Kamideetal:1998}.

The geomagnetic storms can consist of four phases: sudden commencement, initial phase, main phase and recovery phase.
The sudden commencement when it exists, corresponds to the moment when the initial impact of the increased solar wind pressure over the magnetopause occurs.
The initial phase at ground appears as a rapid increase on the H-component over less than 1 h almost simultaneously worldwide.
The main phase of the geomagnetic storm lasts a few hours and is characterized by a decrease in the H-component.
The recovery time corresponds to the gradual increase of the H-component value to its average level.
A detailed description of the morphology of magnetic storms is, for instance, in \cite{Gonzalezetal:1994}.

The intensity of the geomagnetic disturbance in each day is described by indices.
The indices are very useful to provide the global diagnostic of the degree of disturbance level.
There are different indices that can be used depending on the character and the latitude influences in focus.
Considering only the main latitudinal contributions, the ring current dominates at low and middle latitudes and an auroral ionospheric current systems dominates at higher latitudes \citep{Mendesetal:2005}.
Kp, AE and Dst and their derivations are the most used geomagnetic indices.
The Kp index is obtained from the H-component and it is divided in ten levels from 0 to 9 corresponding to the mean value of the disturbance levels within 3-h intervals observed at 13 subauroral magnetic stations \cite[see][]{Bartels1957}.
However, the K index is the most difficult to be physically interpreted due to its variations be caused by any geophysical current system including magnetopause currents, field-aligned currents, and the auroral electrojets \citep{Gonzalezetal:1994}.
The minutely AE index (sometimes $2.5$ minute interval) is also obtained by the H-component measured from magnetic stations (5 to 11 in number) located at auroral zones and widely distributed in longitude.
The AE index provides a measure of the overall horizontal auroral oval current strength.
The index most used in low and mid-latitudes is the Dst index.
It represents the variations of the H-component due to changes of the ring current and is calculated every hour.

The Dst index is described as a measure of the worldwide derivation of the H-component at mid-latitude ground stations from their quiet days values.
At mid-latitude, the H-component is a function of the magnetopause currents, the ring current and tail currents.
\cite{Burtonetal1975} calculated the Dst index as a average of the records from $N$ mid-latitude magnetic stations following,

\begin{equation}
 Dst=\frac{1}{N} \sum^{N}_{i=1}{H_{disturbed}-H_{quiet}}=\overline{\Delta H}
\end{equation}

where $\overline{\Delta H}$ is a local time H average, $H_{disturbed}$ is the H-component measured at disturbed days and $H_{quiet}$, on quiet days.

Other contributions beyond the ring current could be extracted or eliminated with the idea presented by \cite{Burtonetal1975}.
Those authors described the evolution of the ring current by a simple first order differential equation,

\begin{equation}
 \frac{d\,Dst^*}{d\,t}=Q(t)-a\,Dst^*,
\end{equation}

where $Dst^*= Dst-b\,Pdyn^{\frac{1}{2}}+c$. The contribution of the magnetopause currents to $H$ is proportional to the square root of the solar wind dynamic pressure ($Pdyn$), $Q$ represents the injection of particles to the ring current, $a\,Dst^*$ represents the loss of particles with an e-folding time $\frac{1}{a}$ and the constant terms $a$, $b$ and $c$ are determine by the quiet days values of the magnetopause and ring currents.

The Dst index is available on the Kyoto World Data Center at http:// wdc.kugi.kyoto-u.ac.jp/dstdir/index.html.
It is traditionally calculated from four magnetic observatories: Hermanus, Kakioka, Honolulu, and San Juan.
These observatories are located at latitudes below $40^o$ which are sufficiently distant from the auroral electrojets.
The derivation of the Dst index corresponds to three main steps: the removal of the secular variation, the elimination of the Sq variation and the calculation of the hourly equatorial Dst Index (see http://wdc.kugi.kyoto-u.ac.jp/dstdir/dst2/onDstindex.html). 

The traditional method of calculating the baseline for the quiet day variations uses the five quietest day for each month for each magnetic observatory.
In this work, we propose a way to deal with Sq variations by suggesting a method using Principal Component with the wavelet correlation matrix.
This method eliminates the disturbed days using a multiscale process.
Also, we developed an algorithm for extracting the solar quiet variations recorded in the magnetic stations time series, in order words, a  way of estimation of the quiet-time baseline.
To accomplish this task, we separate the solar diurnal variations using hourly data of the H-component using the technique \cite[described in][]{Klausner2011}.
Afterward we applied the principal component wavelet analysis to identify the global patterns in the solar diurnal variations.

The rest of the paper is organized as follows: 
Section~\ref{The Dst index calculation procedure} is devoted to explain the main issues of the Dst index calculation procedure. 
In Section~\ref{Magnetic Data} the analyzed period and data are presented. 
Section~\ref{Methodology} describes the principal component analysis and it is devoted to introduce the suggested method of principal component analysis (PCA) using gapped wavelet transform and wavelet correlation.
It also establishes the identification of the disturbed days. 
The results are discussed in Section~\ref{Results and Discussion}, and finally,
Section \ref{Summary} brings the conclusions of this work.

\section{The Dst index calculation procedure}
\label{The Dst index calculation procedure}

Recently, reconstructing the Dst and removing the quiet-time baseline have been a motivation of several works \citep{KarinenMursula:2005,Mendesetal:2005,KarinenMursula:2006,Mursulaetal:2008,LoveGannon:2009,Klausner2011}.
Despite of these disagreements in the Dst constrution, today, the Dst index remains an important tool in the space weather analysis.

\Citet{KarinenMursula:2005} reconstructed the Dst index (Dxt) following the original formula presented at http://wdc.kugi.kyoto-u.ac.jp /dstdir/dst2/onDstindex.html.
However, they encountered a few issues as: the availability and the data quality , some shifts in the baseline level of the H-component, erroneous data points and some data gaps at all/some magnetic observatories.
The Dst could not be fully reproduced using the original formula because the inadequate information above effects the treatment of the related issues and therefore remains partly unscientific as described by \Citet{KarinenMursula:2005}.

A new corrected and extended version (Dcx) of the Dxt index was proposed by \Citet{KarinenMursula:2006}.
They corrected the Dst index for the excessive seasonal varying quiet-time level which was unrelated to magnetic storms as previously discussed in \Citet{KarinenMursula:2005}.
They also showed that the considerable amount of quiet-time variation is included in the Dxt index but none in the Dcx index.

Another issue related to the derivation of the Dst index is that no treatment is made to normalize the different latitudinal location of the magnetic observatories.
\Citet{Mursulaetal:2008} suggested the normalization of the magnetic disturbances at the four Dst stations with different latitudes by the cosine of the geomagnetic latitude of the respective station.
If no correction is made, they showed for the lowest geomagnetic station, Honolulu, the largest deviations and for the highest station, Hermanus, the lowest deviations of the four station.
The standard deviations reflect the annually averaged effect of the (mainly ring current related) disturbances at each station, \cite[see][for more details]{Mursulaetal:2008}.

\Citet{Mendesetal:2005} evaluated the effect of using more than four magnetic stations and shorter time intervals to calculate the Dst index.
The obtained Dst index profiles using 12, 6 or 4 magnetic station did not show significant discrepancies and the best agreement with the standart Dst was obtained using magnetic stations located at latitudes lower than $35^o$ in both hemispheres.

Although, the increase of symmetrically world-wide distributed magnetic stations did not effect significantly the Dst index, the longitudinal asymmetries of the ring current contributes for the average disturbances of the Four Dst stations be systematically different.
\Citet{Mursulaetal:2010} using an extended network of 17 stations, demonstrated that the local disturbances are ordered according to the station's geographic longitude, where the westernmost station (Honolulu) presented the largest disturbances and contributions to Dst index and the easternmost (Kakioka) the smallest.

\Citet{Klausner2011} studied the characteristics of the Sq variations at a Brazilian station and compared to the features from other magnetic stations to better understand the dynamics of the diurnal variations involved in the monitoring of the Earth's magnetic field.
They used gapped wavelet analysis and the wavelet cross-correlation technique to verify the latitudinal and longitudinal dependence of the diurnal variations.
As previously mentioned by \Citet{Mursulaetal:2010}, \Citet{Klausner2011} also verified that magnetic stations located at lower latitudes and westernmost (Honolulu and San Juan) presented larger correlation to Vassouras than the easternmost stations (as Kakioka).

Some important aspects for the construction of the Dst index as described by \Citet{LoveGannon:2009} are: the utilization of the original data, the inspection in time and frequency domains (removal of diurnal variation) and the consideration of the distinction between stationary and non-stationary time series ingredients which applies to the secular variation. 
Also as mentioned by \Citet{LoveGannon:2009}, some patterns of the global magnetic disturbance field are well understood and some are not which means that there is still a lot to learn about the magnetosphere, magnetic storm and Earth-Sun relationship.

\section{Magnetic Data}
\label{Magnetic Data}

In this paper, we use ground magnetic measurements to estimate the quiet-time baseline.
We select the four magnetic observatories used to calculate the Dst index: Hermanus (HER), Kakioka (KAK), Honolulu (HON), and San Juan (SJG), plus other $9$
different magnetic observatories reasonably homogeneously distributed world wide.
One of these nine chosen stations is Vassouras (VSS) located under the South Atlantic Magnetic Anomaly (minimum of the geomagnetic field intensity).
The geomagnetic data use in this work relied on data collections provided by the INTERMAGNET programme (http://www.intermagnet.org).

\begin{figure}[ht]
  \centering
    \includegraphics[width=1.0\textwidth]{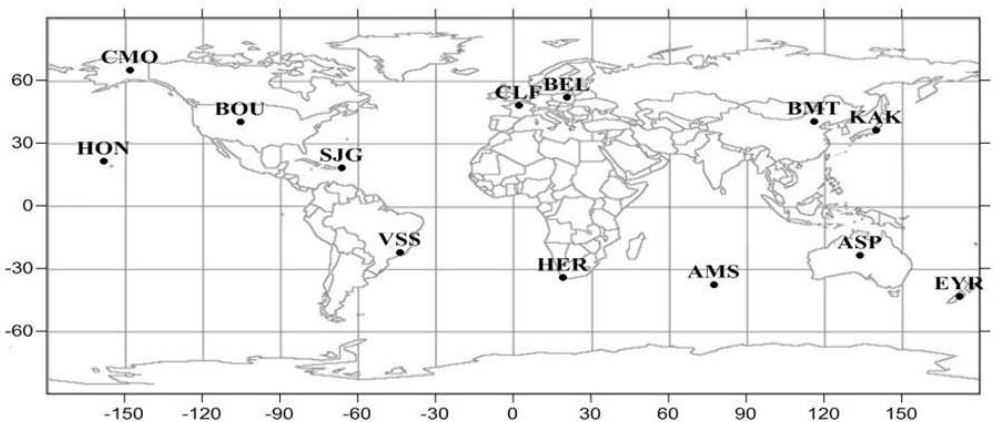}
  \caption{Geographical localization of the stations used in this work and their respective IAGA code.}
  \label{fig:MapStations}
\end{figure}

The distribution of the magnetic stations, with their IAGA codes, is given in Fig.~\ref{fig:MapStations}. 
The corresponding codes and locations are given in Table~\ref{table:ABBcode}. 
These selection of magnetic stations correspond to the same selection used in a previous work o\citep{Klausner2011} for the same reasons (exclusion of the major influence of the auroral and equatorial electrojets).
In this work, we only use the data interval corresponding to the year 2007.
We also apply the same methodology to identify geomagnetically quiet days used by  \Citet{Klausner2011}.
We consider quiet days, only those days in which the Kp index is not higher than 3+.

As at low latitudes the horizontal component (H) is mostly affected by the intensity of the ring current, we decided to use only the hourly mean value series of this component.
The magnetic stations present available data in Cartesian components (XYZ system).
The Conversion to horizontal-polar components (HDZ system) is very simple \citep[see][for more details]{Campbell1997}.
The system's conversion was performed in all the chosen magnetic stations.

\begin{table}[ht]
 \caption{INTERMAGNET network of geomagnetic stations used in this study.}
\centering
\begin{tabular}{c c c c c }
\hline
Station & \multicolumn{2}{c}{Geographic coord.} & \multicolumn{2}{c}{Geomagnetic coord.}\\
\cline{2-5}
          & Lat.($^o$) & Long.($^o$) & Lat.($^o$) & Long.($^o$)  \\[0.5ex]
\hline
AMS     &-37.83     &77.56         &-46.07   &144.94  \\
ASP     &-23.76     &133.88        &-32.50   &-151.45   \\
BEL     &51.83      &20.80         &50.05    &105.18   \\
BMT     &40.30      &116.20        &30.22     &-172.55   \\
BOU     & 40.13     & -105.23      & 48.05    &-38.67   \\
CLF     &48.02       &2.26         &49.56      &85.72   \\
CMO     &64.87      &-147.86       &65.36      &-97.23   \\
EYR     &-43.42     &172.35        &-46.79     &-106.06   \\
HER     &-34.41     &19.23         &-33.89     &84.68   \\
HON     &21.32      &-158.00       &21.59      &-89.70   \\
KAK     & 36.23     & 140.18       & 27.46     &-150.78\\
SJG     &18.12      &-66.15        &27.93      &6.53  \\
VSS     & -22.40    & -43.65       &-13.43     &27.06 \\[1ex]
\hline
\end{tabular}
\\Source: http://wdc.kugi.kyoto-u.ac.jp/igrf/gggm/index.html (2010)
\label{table:ABBcode}
\end{table}

\section{Methodology}
\label{Methodology}

The method used in this study is based on the principal component analysis (PCA) using gapped wavelet transform and wavelet correlation to characterize the global diurnal variation behavior.
To identify periods of magnetic disturbance, we use the discrete wavelet transform.
This technique is employed to analyze the removal of disturbed days from the magnetograms, and consequently, from the reconstructed Sq signal. 
Also in this section, a combined methodology using the PCA and gapped wavelet transform is briefly described.
Following, we present an identification method to distinguish the disturbed days using discrete wavelet coefficients.

\subsection*{Global geomagnetic behavior analysis}

Among the several available methods of analysis, PCA is a particularly useful tool in studying large quantities of multi-variate data.
PCA is used to decompose a time-series into its orthogonal component modes, the first of which can be used to describe the dominant patterns of variance in the time series \cite{Murray1984}.
The PCA is able also to reduce the original data set of two or more observed variables by identifying the significant information from the data.
Principal Components (PCs) are derived as the eigenvectors of the correlation matrix between the variables.
Their forms depend directly on the interrelationships existing within the data itself.
The first PC is a linear combination of the original variables, which when used as a linear predictor of these variables, explains the largest fraction of the total variance. The second, third PC, etc., explain the largest parts of the remaining variance \cite{Murray1984}.

As explained by \Citet{Yamada2002}, the interpretation of the eigenvectors and the eigenvalues can be described as follow, the eigenvectors are the normalized orthogonal basis in phase space, and also, the set of vectors of the new coordinate system in the space, different from the coordinate system of the original variables; the eigenvalues are the corresponding variance of the distribution of the projections in the new basis.

In order to isolate the global contributions of each PCs mode, we applied PCA using the wavelet correlation matrix computed by gapped wavelet transform.
This wavelet correlation matrix was introduced in \cite{Nesme-Ribesetall1995}.
We joined the properties of the PCA, which are the compression of large databases and the simplification by the PCs modes, and properties of the wavelet correlation matrix, which is the correlation at a given scale, $a$, in this case, the scale corresponded to the pseudo-period of $24$ hours.

\subsection*{Identification of magnetic disturbance}

The wavelet analysis has the following propriety: the larger amplitudes of the wavelet coefficients are associated with locally abrupt signal changes or ``details'' of higher frequency. 
In the work of \Citet{MendesMag2005} and the following work of \cite{MendesdaCostaetal:2011}, a method for the detection of the transition region and the exactly location of this discontinuities due to geomagnetic storms was implemented. 
In these cases, the highest amplitudes of the wavelet coefficients indicate the singularities on the geomagnetic signal in association with the disturbed periods.
On the other hand,  when the magnetosphere is under quiet conditions for the geomagnetic signal, the wavelet coefficients show very small amplitudes.
In this work, we applied this methodology with Daubechies orthogonal wavelet function of order 2 on the one minute time resolution with the pseudo-periods of the first three levels of 3, 6 and 12 minutes.

\section{Results and Discussion}
\label{Results and Discussion}

In this section, we will present the results of reconstructed baseline for the global quiet days variation using PCA technique implemented with gapped wavelet transform and wavelet correlation.
Also, we will apply DWT to evaluate the day-by-day level of geomagnetic disturbance using KAK magnetic station as reference.

\begin{figure}[ht]
\centering
     \includegraphics[width=14cm]{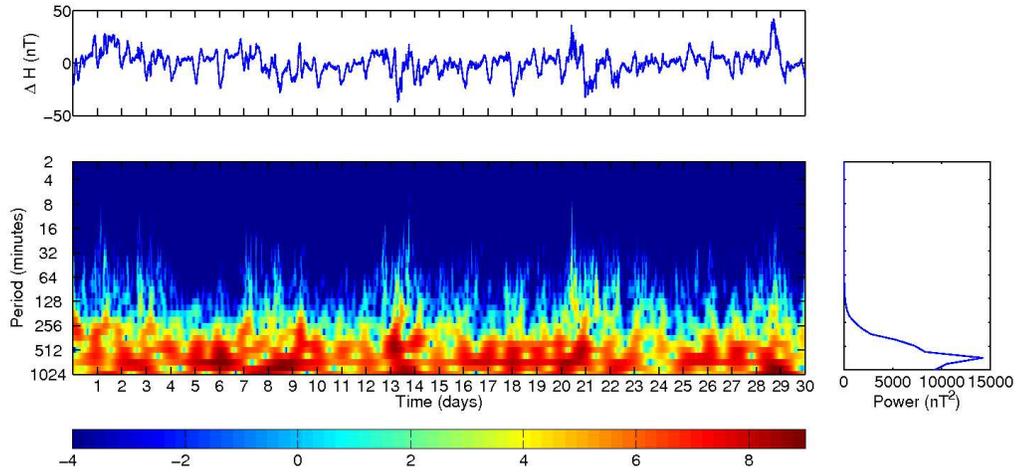}\\
\caption{The GWT analysis of KAK magnetogram. At top, it shows the H-component of KAK at June, 2007 used 
  for the wavelet analysis, at bottom left, the scalogram using Morlet wavelet, logarithmic scaled 
  representing $\log2{(|W(a,b)|)}$, and at bottom right, the global wavelet spectrum.}
\label{fig:KAKescalogram}
\end{figure}

Fig.~\ref{fig:KAKescalogram} shows an example of the geomagnetic behavior presented at the June, 2007 magnetogram of Kakioka using continuous gapped wavelet transform (GWT).
The GWT can be used in the analysis of non-stationary signal to obtain information on the frequency or scale variations and to detect its structures localization in time and/or in space \citep[see][for more details]{Klausner2011}.
It is possible to analyze a signal in a time-scale plane, called so the wavelet scalogram.
In analogy with the Fourier analysis, the square modulus of the wavelet coefficient, $|W(a,b)|^2$, is used to provide the energy distribution in the time-scale plane. 
In the GWT analysis, we can also explore the central frequencies of the time series through the global wavelet spectrum which is the variance average at each scale over the whole time series, to compare the spectral power at different scales.
This figure shows the H-component (top), the wavelet square modulus (bottom left) and the global wavelet spectrum (total energy in each scale -- bottom right).
In the scalogram, areas of stronger wavelet power are shown in dark red on a plot of time (horizontally) and time scale (vertically).
The areas of low wavelet power are shown in dark blue.

In Fig.~\ref{fig:KAKescalogram}, it is possible to notice peaks of wavelet power on the scalogram at the time scale corresponding to $8$ to $16$ minutes of period.
This periods are associated to PC5 pulsations during disturbed periods.
Also, it is possible to notice a maximum of wavelet power at the time scale corresponding to harmonic periods of the 24 hours such as 6, 8, 12 hours.
Those periods are related to the diurnal variations.

The GWT technique is able to analyze all the informations present on the magnetograms.
It is an auxiliary tool to localize on time/space the PC1--PC5 pulsations \citep{Saito1969}.
However, the scalogram provides a very redundancy information which difficult the analysis of each decoupling phenomenum.
For that reason, we preferred to use DWT to evaluate the day-by-day level of geomagnetic disturbance.

\begin{figure}[ht]
\centering
     \includegraphics[width=14cm]{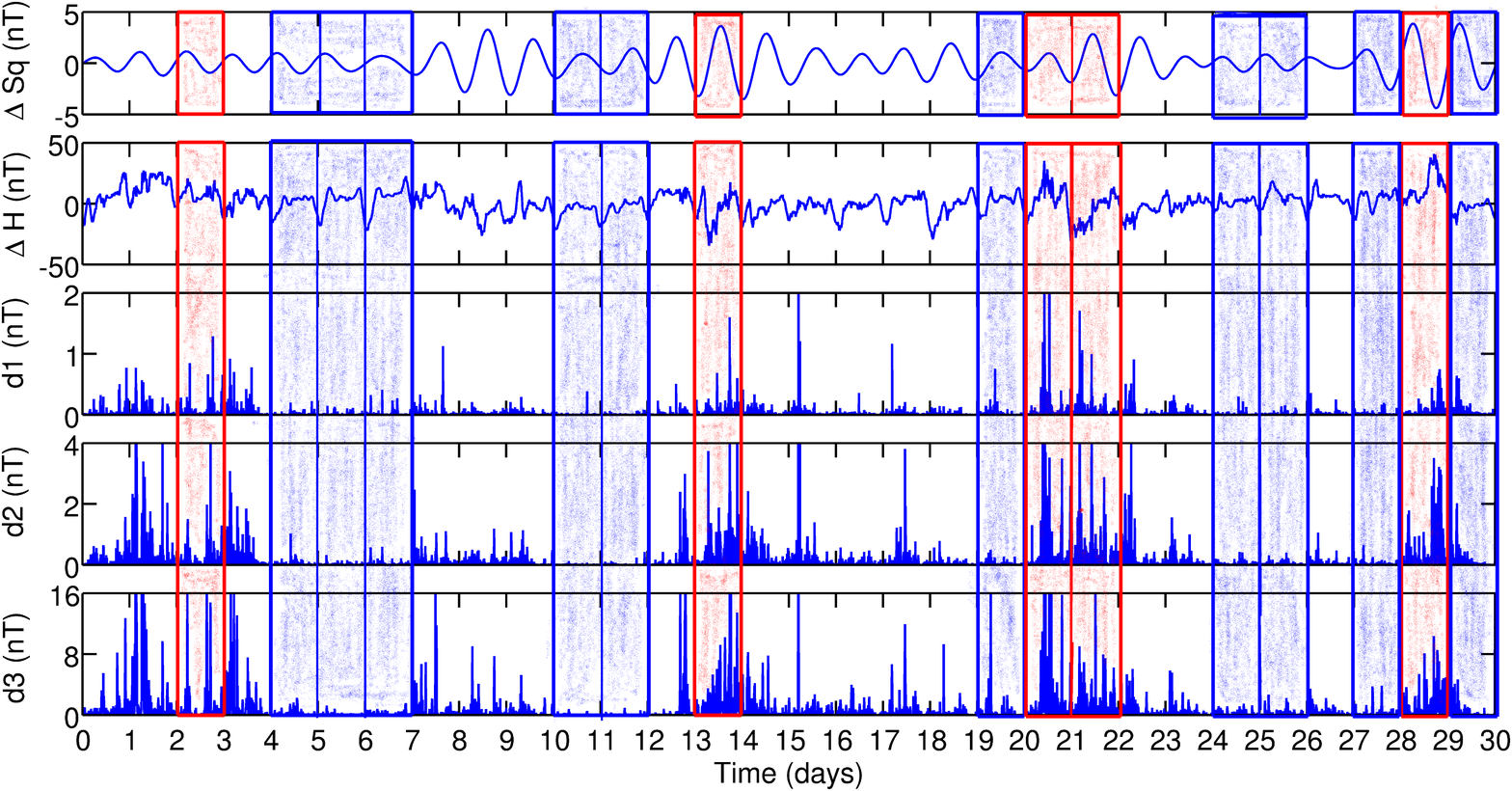}\\
\caption{The amplitude variation of the reconstructed Sq signal with the highlighted $10$ geomagnetically quietest days (blue) and $5$ most disturbed days (red), the H-component average variation for KAK obtained at June, 2007 and the square root wavelet coefficients amplitudes $d1$, $d2$ and $d3$ with the pseudo-periods of 3, 6 and 12 minutes.}
\label{fig:SqJun}
\end{figure}

Fig.~\ref{fig:SqJun} is composed of two graphs, the first one presents the reconstructed line of the Sq variation where the first $10$ geomagnetically quietest days of each month are highlighted in blue and the $5$ most disturbed days in red and the second one presents the discrete wavelet analysis of the geomagnetic horizontal component obtained at Kakioka station, Japan.

Using our criteria of removing disturbed days, we consider as gaps of 3rd, 14th, 21th and 29th day.
In our case, the gapped wavelet technique is very helpful because it reduces two effects: the presence of gaps and the boundary effects due to the finite length of the data, for more details see \Citet{Klausner2011}.

The first graph shows the amplitude range between $-5$ and $5$ nT and presents a complex pattern.
It is possible to notice that the larger amplitudes of the reconstructed Sq signal correspond to the periods between the days 8--10, 13--15, 21--23 and 29--30.
These periods correspond to the disturbed days.

Table~\ref{table:Sq10calmos} shows these $10$ quietest days and $5$ most disturbed days of each month set by the \textit{GeoForschungsZentrum (GFZ) Potsdam} through the analysis of the Kp index that are highlighted on the second graph.
The year of $2007$ is a representative year of minimum solar activity and it is used in this analysis due to have less disturbed periods (see our considerations in Section~\ref{Magnetic Data}).
By analyzing these highlighted days, we expect to find out if there is a correlation between the days classified as quiet days and a small Sq amplitude variation.

\begin{table}[ht]
\scriptsize{
 \caption{The first $10$ geomagnetically quietest days and first $5$ most disturbed days set by the \textit{GeoForschungsZentrum (GFZ) Potsdam}}
\centering
\begin{tabular}{c c c c c c c c c c c c c c c c }
\hline
Month/year & \multicolumn{10}{c}{$10$ quietest days}& \multicolumn{5}{c}{$5$ most disturbed days}\\
\cline{1-16}	&	q1	&	q2	&	q3	&	q4	&	q5	&	q6	&	q7	&	q8	&	q9	&	q10	&	d1	&	d2	&	d3	&	d4	&	d5	\\
\hline
Mar	&	20	&	21	&	3	&	19	&	9	&	29	&	22	&	18	&	31	&	4	&	13	&	24	&	6	&	7	&	14	\\
Jun	&	5	&	12	&	6	&	7	&	11	&	26	&	25	&	20	&	30	&	28	&	14	&	21	&	22	&	3	&	29	\\
Sep	&	13	&	9	&	10	&	11	&	12	&	16	&	17	&	19	&	26	&	15	&	29	&	2	&	28	&	23	&	27	\\
Dec	&	3	&	8	&	4	&	25	&	7	&	26	&	2	&	29	&	15	&	6	&	18	&	17	&	11	&	20	&	21	\\
\hline
\end{tabular}
SOURCE: http://wdc.kugi.kyoto-u.ac.jp/qddays/index.html.\\
}
\label{table:Sq10calmos}
\end{table}

The second graph shows the discrete wavelet analysis applied to geomagnetic minutely signal from KAK using Daubechies orthogonal wavelet family 2.
From top to bottom in this graph, the H-component of the geomagnetic field and the first three levels of the square wavelet coefficients denoted by d1, d2 and d3.
This analysis uses the methodology developed by \Citet{MendesMag2005}, and posteriorly applied by \citet{MendesdaCostaetal:2011} and \Citet{Klausner2011b}.

\begin{figure}[ht]
\centering
\psfrag{Sum Coeff. Wav.}[c][][0.8]{$\;\;EWC$}
\psfrag{Time (days)}[c][][0.8]{$Time\;(days)$}
  \includegraphics[width=14cm]{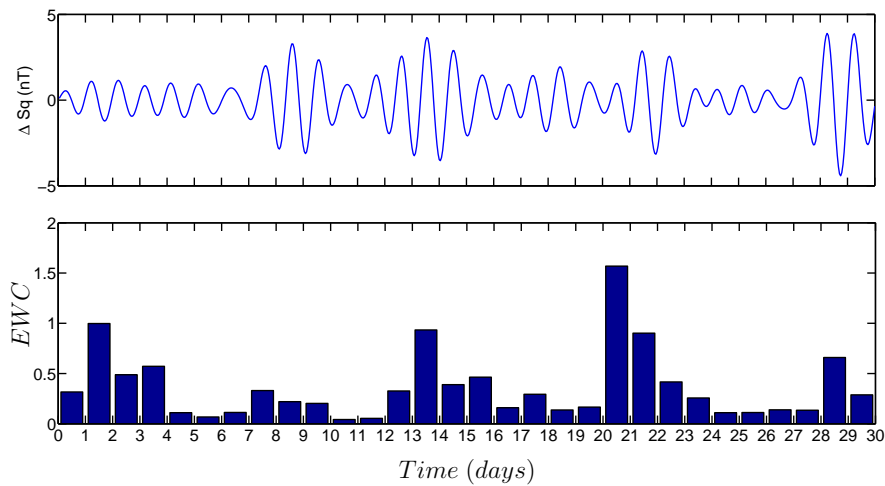}\\
\caption{The comparative of the global Sq behavior and the effectiveness wavelet coefficients for the month of June, 2007.}
\label{fig:SUMSqJun}
\end{figure}

In order to facilitate the evaluation of the quiet periods obtained by the discrete wavelet analysis applied to geomagnetic signal from KAK, we also developed a methodology (effectiveness wavelet coefficients (EWC)) to interpret the results shown in Fig.~\ref{fig:SUMSqJun}.
The EWC corresponds to the weighted geometric mean of the square wavelet coefficients per hour.
It is accomplished by weighting the square wavelet coefficients means in each level of decomposition as following

\begin{equation}
 EWC=\frac{4\,\sum_{i=1}^{N}d1+2\,\sum_{i=1}^{N}d2+\sum_{i=1}^{N}d3}{7},
\end{equation}

where $N$ is equal to $60$ because our time series has one minute resolution.

Through Fig.~\ref{fig:SUMSqJun}, it is possible to compare the global Sq behavior (top graph) with the analysis of quiet and disturbed days obtained by one representative magnetic station of medium/low latitudes (KAK -- bottom graph) in order to verify situations in which the global Sq behavior presents less or more variability.
This analysis allows us to validate the quietest days, and evaluate the most disturbed days in order to establish a reliable method of global Sq analysis obtained from medium/low latitude magnetic stations influenced only by the ionosphere.

In Fig.~\ref{fig:SUMSqJun}, the global Sq behavior (top graph) shows larger amplitudes during the periods between the days 8--10, 12--16, 22--23 and 29--30.
Most of these periods correspond to the days where the EWCs have an increase of their values as shown KAK analysis (bottom graph).
The increase of the EWCs values occurs during the periods between the days 1--4, 8--10, 13--18, 21--24 and 29--30.
The EWCs can help us also to interpret the results obtained in each day, and can help us to evaluate the quietest and most disturbed days measure by the selected magnetic station of medium/low latitudes, KAK, as shown in Table~\ref{table:Sq10calmosDWT}.

\begin{table}[ht]
\scriptsize{
 \caption{The $10$ geomagnetically quietest days and $5$ most disturbed days obtained by the discrete wavelet analysis}
\centering
\begin{tabular}{c c c c c c c c c c c c c c c c }
\hline
Month/year & \multicolumn{10}{c}{$10$ quietest days}& \multicolumn{5}{c}{$5$ most disturbed days}\\
\cline{1-16}	&	q1	&	q2	&	q3	&	q4	&	q5	&	q6	&	q7	&	q8	&	q9	&	q10	&	d1	&	d2	&	d3	&	d4	&	d5	\\
\hline
Mar	&	3	&	10	&	20	&	9	&	21	&	8	&	2	&	19	&	22	&	4	&	13	&	25	&	12	&	15	&	24	\\
Jun	&	11	&	12	&	6	&	25	&	5	&	26	&	7	&	28	&	19	&	27	&	21	&	2	&	14	&	22	&	29	\\
Sep	&	12	&	11	&	10	&	9	&	17	&	18	&	16	&	13	&	26	&	19	&	27	&	22	&	29	&	2	&	20	\\
Dec	&	25	&	7	&	3	&	8	&	1	&	24	&	4	&	6	&	16	&	28	&	17	&	18	&	11	&	20	&	10	\\
\hline
\end{tabular}
}
\label{table:Sq10calmosDWT}
\end{table}

The same methodology and analysis comparing the global Sq behavior and the EWCs from KAK is done for the month of March, September and December,2007.

\begin{figure}[ht]
\centering
\psfrag{Sum Coeff. Wav.}[c][][0.8]{$\;\;EWC$}
\psfrag{Time (days)}[c][][0.8]{$Time\;(days)$}
   \includegraphics[width=14cm]{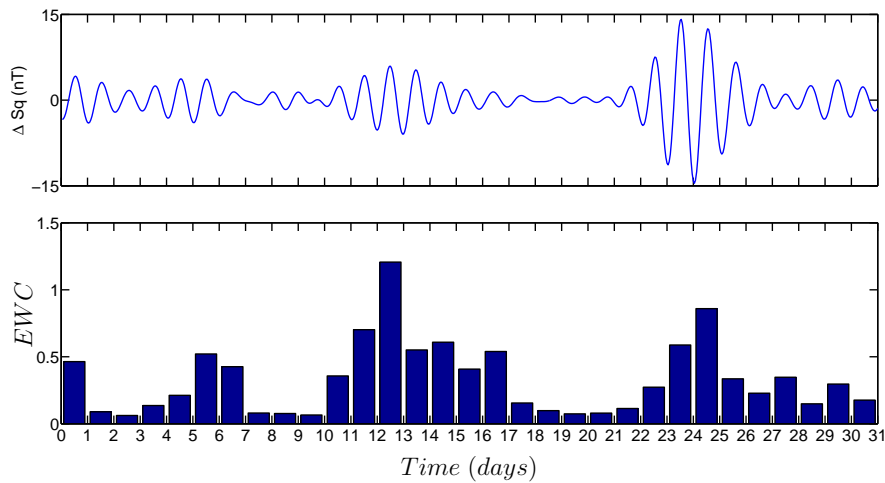}\\
  \caption{The comparative of the global Sq behavior and the effectiveness wavelet coefficients for the month of March, 2007}
\label{fig:SUMSqMar}
\end{figure}

Fig.~\ref{fig:SUMSqMar} shows the comparative of the global Sq behavior and the EWCs from KAK for the month of March, 2007.
The amplitude range of global Sq signal is between $-15$ and $15$ nT, and, it also shows a complex pattern.
It is possible to notice that the larger amplitudes of the reconstructed Sq signal correspond to the periods between the days 6--7, 11--17 and 23--28.
Once more, these periods correspond to the disturbed days.
The increase of the EWCs values occurs during the periods between the days 1, 6--7, 11--17, 23--28 and 30--31.
We can notice that the increase of the amplitude of the reconstructed Sq signal correspond to the increase of the EWCs magnitude.

\begin{figure}[ht]
\centering
\psfrag{Sum Coeff. Wav.}[c][][0.8]{$\;\;EWC$}
\psfrag{Time (days)}[c][][0.8]{$Time\;(days)$}
\includegraphics[width=14cm]{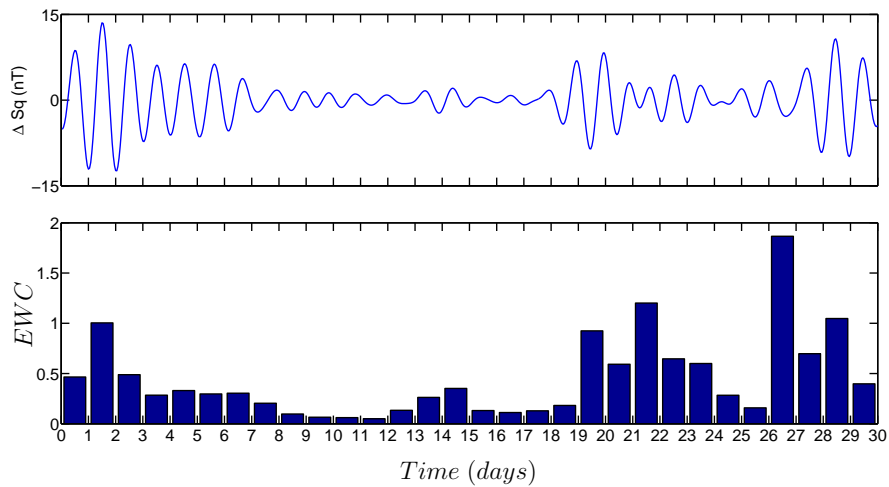}\\
\caption{The comparative of the global Sq behavior and the effectiveness wavelet coefficients for the month of September, 2007}
\label{fig:SUMSqSet}
\end{figure}

The amplitude range for September, 2007, is between $-15$ and $15$ nT and the larger amplitudes of the reconstructed Sq signal correspond to the periods between the days 1--7, 19--23 and 26--30, see Fig,~\ref{fig:SUMSqSet}.
The increase of the EWCs values occurs during the periods between the days 1--3, 20--24 and 27--30.
Comparing these two analysis, the global and the EWCs, we verify that the KAK magnetic behavior represents well the increase of global Sq oscillations.

\begin{figure}[ht]
\centering
\psfrag{Sum Coeff. Wav.}[c][][0.8]{$\;\;EWC$}
\psfrag{Time (days)}[c][][0.8]{$Time\;(days)$}
  \includegraphics[width=14cm]{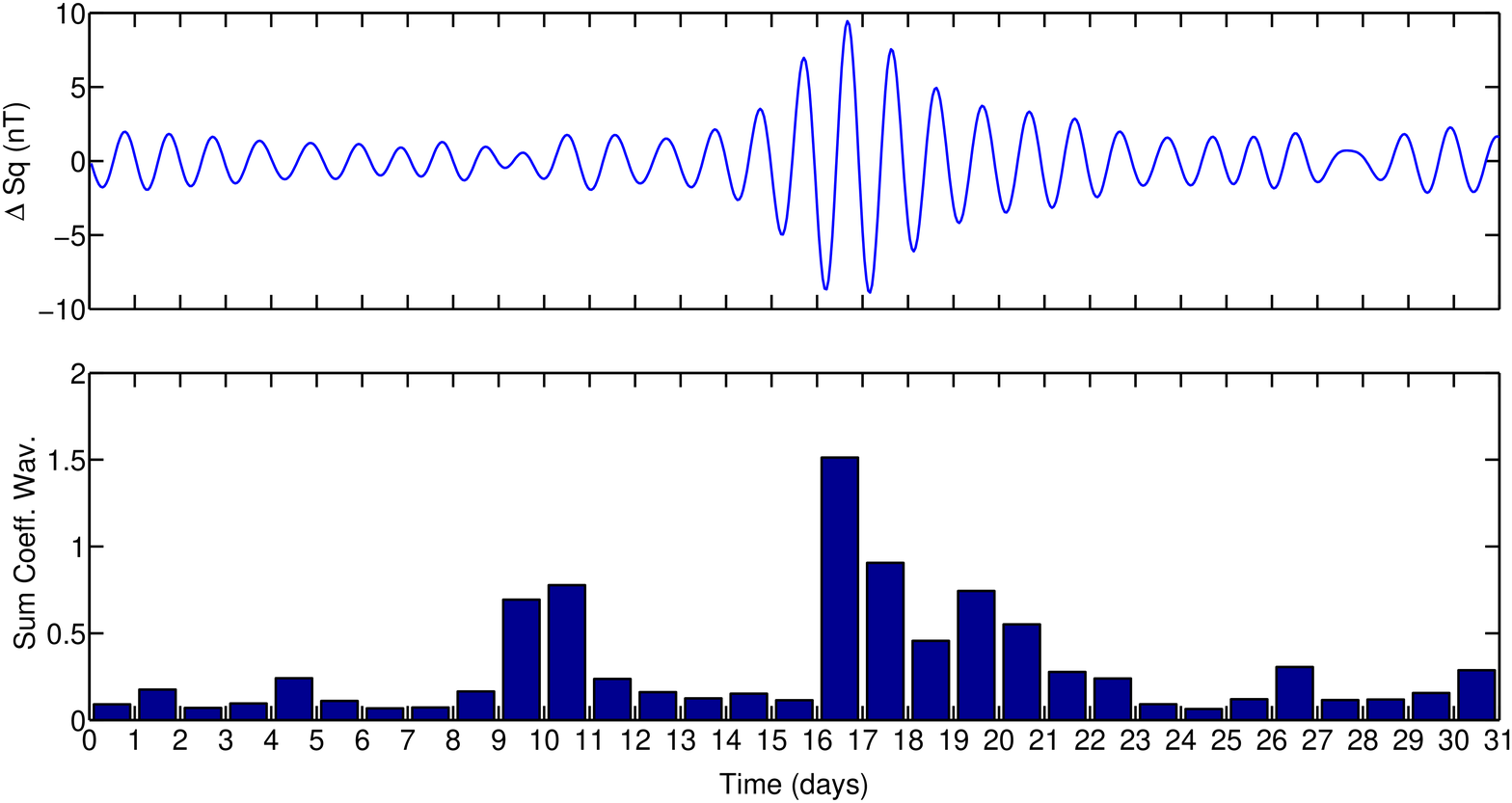}\\
\caption{The comparative of the global Sq behavior and the effectiveness wavelet coefficients for the month of December, 2007}
\label{fig:SUMSqDec}
\end{figure}

In Fig.~\ref{fig:SUMSqDec}, the amplitude range is between  $-10$ and $10$ nT and the larger amplitudes correspond to the periods between the days 15--23.
Also, the increase of the EWCs values occur during the periods between the days 10--11 and 17--21.
Once more, when we compare these two analysis, the global and the EWCs, we verify that the KAK magnetic behavior represents well the increase of global Sq oscillations.

We observe in Figs.~\ref{fig:SUMSqJun}, \ref{fig:SUMSqMar}, \ref{fig:SUMSqSet} and \ref{fig:SUMSqDec}, in most of the cases, the major amplitude fluctuations of the reconstructed Sq signal correspond to the most disturbed days and minor fluctuations, to the quietest days.
However, the diurnal global variability shows a complexity on the amplitude variation pattern even during geomagnetically quiet periods.
Through this study we compare the amplitude variation of the reconstructed Sq signal to effectiveness wavelet coefficients obtained at KAK with the purpose of understanding the complexity of the diurnal global variability.

\section{Conclusions}
\label{Summary}

In this work, we suggest an alternative approach for the calculation of the Sq baseline using wavelet and PCA techniques.
This new approach address some issues, such as, the availability and the quality of data, abrupt changes in the level of the H-component, erroneous points in the database and the presence of gaps in almost all the magnetic observatories.
To fulfill this purpose, we reconstruct the Sq baseline using the wavelet correlation matrix with scale of $24$ hours (pseudo-period).
The PCA/wavelet method uses the global variation of first PCA mode that also corresponds to phenomena with periods of $24$ hours.
This study shows that the largest amplitude oscillation of the reconstructed signal (Sq baseline) corresponded to the most disturbed days and the smaller oscillations to the quietest days.
This result is consistent with the expected Sq variations.

\section{Acknowledgments}
V. Klausner wishes to thanks CAPES for the financial support of her PhD (CAPES -- grants 465/2008) and her Postdoctoral research (FAPESP -- 2011/20588-7).
This work was supported by CNPq (grants 309017/2007-6, 486165/2006-0, 308680/2007-3, 478707/2003, 477819/2003-6, 382465/01-6), FAPESP (grants
2007/07723-7) and CAPES (grants 86/2010-29, 0880/08-6, 86/2010-29, 551006/2011-0, 17002/2012-8). 
Also, the authors would like to thank the INTERMAGNET programme for the datasets used in this work.


%
%


\bibliographystyle{abbrv}

\bibliographystyle{elsarticle-harv}





\end{document}